# Status of DUNE Offline Computing

*Michael* Kirby[1,][*] on behalf of the DUNE Collaboration

[1]Fermi National Accelerator Laboratory

**Abstract.** We summarize the status of Deep Underground Neutrino Experiment (DUNE) Offline Software and Computing program. We describe plans for the computing infrastructure needed to acquire, catalog, reconstruct, simulate and analyze the data from the DUNE experiment and its prototypes in pursuit of the experiment's physics goals of precision measurements of neutrino oscillation parameters, detection of astrophysical neutrinos, measurement of neutrino interaction properties and searches for physics beyond the Standard Model. In contrast to traditional HEP computational problems, DUNE's Liquid Argon Time Projection Chamber data consist of simple but very large (many GB) data objects which share many characteristics with astrophysical images. We have successfully reconstructed and simulated data from 4% prototype detector runs at CERN. The data volume from the full DUNE detector, when it starts commissioning late in this decade will present memory management challenges in conventional processing but significant opportunities to use advances in machine learning and pattern recognition as a frontier user of High Performance Computing facilities capable of massively parallel processing. Our goal is to develop infrastructure resources that are flexible and accessible enough to support creative software solutions as HEP computing evolves.

## 1 Introduction

The Deep Underground Neutrino Experiment (DUNE) is the flagship experiment at the Fermi National Accelerator Laboratory and is pursuing a broad range of physics measurements over the next few decades. DUNE plans to utilize four large liquid argon time projection chamber (LAr TPC) detector modules (two in the preliminary phase, and two additional in the second phase of the experiment) at the Sanford Underground Research Facility (SURF). The far detector complex will be used in conjunction with a multipurpose near detector facility and the high intensity neutrino beam from the Long-Baseline Neutrino Facility (LBNF), both located at FNAL, to make high precision measurements of neutrino oscillations parameters. In addition to measurements of neutrino oscillation parameters (e.g. neutrino mass ordering, CP-violating phase, and PMNS unitarity), the physics program for DUNE includes searches for physics beyond the standard model, supernova burst neutrinos, solar neutrino measurements, and more. The design of the far detector LAr TPC technology is currently being prototyped with the ProtoDUNE Horizontal Drift and Vertical Drift cryostats located at the CERN EHN1 Neutrino Platform. To meet the needs of the experiment, the DUNE Computing Consortium is actively supporting these prototype activities while designing and constructing infrastructure, middleware, and systems to support the full physics program of DUNE throughout the next several decades.

---

[*]e-mail: kirby@fnal.gov

## 2 Computing Model Strategic Design

The DUNE Offline Software and Computing (S&C) program has made several choices to guide its strategic approach for meeting the resource needs and software requirements of the DUNE physics program. The primary software strategy has been the adoption of community developed software and services where appropriate (Rucio[1] for replica management, GlideInWMS[3] for distributed computing workload management, etc). While common solutions carry the benefit of long-term stability and support, there are computing requirements that are unique to DUNE from both the physics program and detector design that must be addressed. One example is the unique requirements for the reconstruction of dynamic readout time windows for supernova burst searches in the far detector which can be as large as 160 TB of data from 100 seconds of single-module, continuous readout. The identification of these unique requirements has been a focal point of the Computing Consortium and they are documented in the DUNE Offline Computing Conceptual Design Report[4].

The DUNE Computing Consortium has chosen a computing resource model based upon the HSF DOMA working group proposal [2] and consists of a flatter architecture than the current tiered WLCG computing model. Within the DUNE resource model, there are three primary resources that can be provided as a service: processing (CPU, GPU, HPC, etc), disk storage, and tape archival storage. An institution can contribute one or multiple services above a minimum size for each ( 0.5 - 1 PB of disk storage, minimum RAM/core). An institution must also provide networking and support that is commensurate with the amount and number of services provided. This model takes advantage of existing distributed computing sites that can add additional resources that are specific to the needs of DUNE without requiring a predetermined combination of services. As well, this model allows 'processing-only' sites without permanent storage allocations that will stream data from nearby storage elements. Figure 1 shows a diagram of the conceptual design of the DUNE Computing Resource Model showing logical location of SURF, LBNF, and EHN1 along with several types of resources sites.

## 3 DUNE Computing Resource Estimates

In order to appropriately plan for future computing resource needs, the DUNE Computing Consortium has developed a python-based software package that projects that annual needs for processing, disk storage, and archival tape storage. The software takes into account anticipated trigger rates from LBNF beam interactions, solar neutrinos, cosmic ray backgrounds, hypothetical supernova-trigger rate, and calibrations from the far detector modules and near detector facility. The projected data rates for a single horizontal drift far detector module are shown in Table 1. To provide fast and efficient access to data, DUNE has established policies for the placement of both raw and derived datasets. To both ensure no loss of raw data and efficient processing, two copies of the raw data will be stored on archival tape with one copy at Fermilab and the second copy stored abroad. Derived and simulation data will have a single copy on tape archive, but two copies on disk while those datasets are active. Strict retention and removal policies for obsolete derived datasets from disk storage are also included as part of the resource model. Rucio and CERN FTS3 [5] have been successfully deployed to execute this replica policy across more than a dozen storage elements around the world. Using this replica policy and estimates for prototypes, near detector, and simulation campaigns, an estimate for total resources needs projecting out to 2040 have been calculated and the estimates for disk and tape storage are shown in Figure 2.

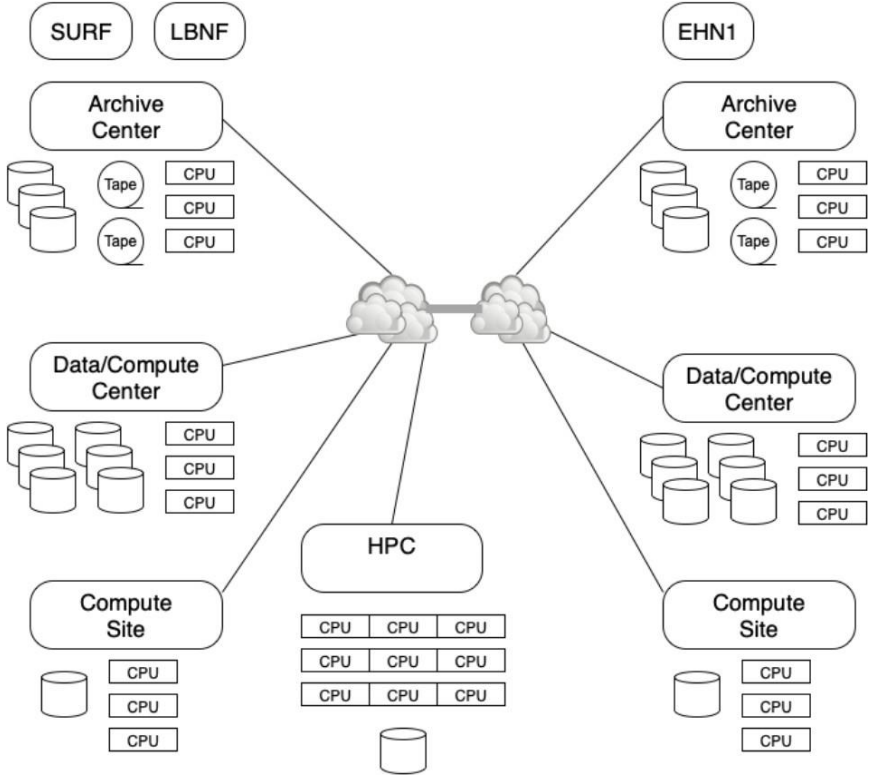

**Figure 1.** DUNE Computing Resource Model diagram including logical representation of SURF, LBNF, and EHN1 as locations for experiment or prototype detectors and several different types of resources sites and the services that those sites provide.

**Table 1.** Data sizes and rates for different processes in each horizontal drift detector module. Uncompressed data sizes are given. As readouts will be self-triggering, a 2.6 ms drift-window readout time is used. We assume beam uptime of 50% and 100% uptime for non-beam science. [4]

| Process | Rate/module | size/instance | size/module/year |
|---|---|---|---|
| Beam event | 41/day | 3.8 GB | 30 TB/year |
| Cosmic rays | 4,500/day | 3.8 GB | 6.2 PB/year |
| Supernova trigger | 1/month | 140 TB | 1.7 PB/year |
| Solar neutrinos | 10,000/year | ≤3.8 GB | 35 TB/year |
| Calibrations | 2/year | 750 TB | 1.5 PB/year |
| Total | | | 9.4 PB/year |

## 4 DUNE Data Challenge Summer 2022

In order to test the infrastructure and reliability of DUNE computing resources, the Computing Consortium engaged in a data challenge during the summer of 2022. The goal of the data challenge was to test all services and procedures needed during the operations of ProtoDUNE Horizontal Drift (HD) and Vertical Drift (VD) prototypes in 2023 and 2024. Additionally, the data challenge tested access and efficiency of distributed resources from more than 20 sites around the world. The data challenge was conducted in two phases. The first phase was focused on the ProtoDUNE HD/VD raw data pipeline and consisted of testing the data path with synthetic data files from the CERN Neutrino Platform, transferred to CERN EOS storage, and then onward to storage elements at FNAL. Additionally, the data files were registered with Rucio, and replicas automatically produced at RAL and IN2P3 storage elements based upon formal data policies. This phase was successful and achieved throughput greater than the estimated ProtoDUNE raw data rate. The data path of the raw data for ProtoDUNE HD/VD is shown in Figure 3. The second phase of the data challenge involved successful scheduling and running of distributed processing jobs that would consume the fake raw data used in initial phase of the data challenge. The workflow was successfully controlled by the justIN workflow system that provides late binding between distributed job slots and task, and was able to sustain 5000 concurrent jobs needed for keep-up processing of the ProtoDUNE data.

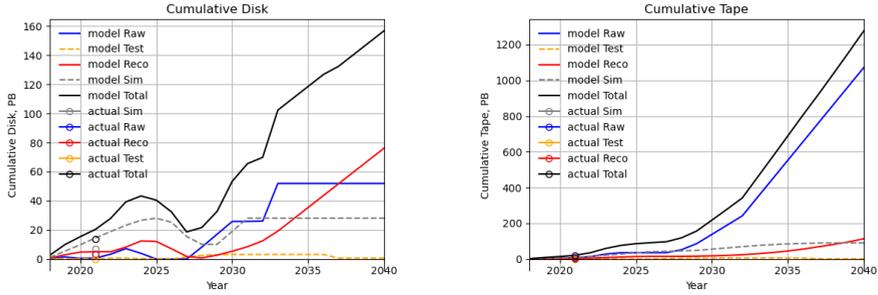

**Figure 2.** DUNE Computing Resource Model projections for needed storage volumes on both disk and tape resources needs in PB out until 2040. This estimate includes retention policies, multiple replicas, and deletion of obsolete derived datasets.

## 5 DUNE Offline Computing Timeline

The DUNE Offline Computing activities are driven in response to both current operational needs and for the long-term development of software and infrastructure that will meet the needs of the full DUNE experiment: far detector, near detector, simulations, and physics analyses. The timeline shown in Figure 4 shows the current preliminary timeline for computing operations and development along with commissioning and operating dates for experiment prototypes and detectors. The Data Challenge discuss in Section 4 was done in preparation for the operation of ProtoDUNE HD and VD in 2023 and 2024 which is expected to produce the largest amount of raw data of all prototype detectors and the largest amount of processing. Following the operation of ProtoDUNE HD and VD, DUNE will also participate in the WLCG Data Challenge 2024 with the unique configuration that raw data will be produced in North America and then distributed to storage elements in Europe and elsewhere. The initial goal is to distribute 25% of the raw data rate for the DUNE far detector (2 Gbit/s) around the globe for storage and subsequent processing. Additionally, job submission and

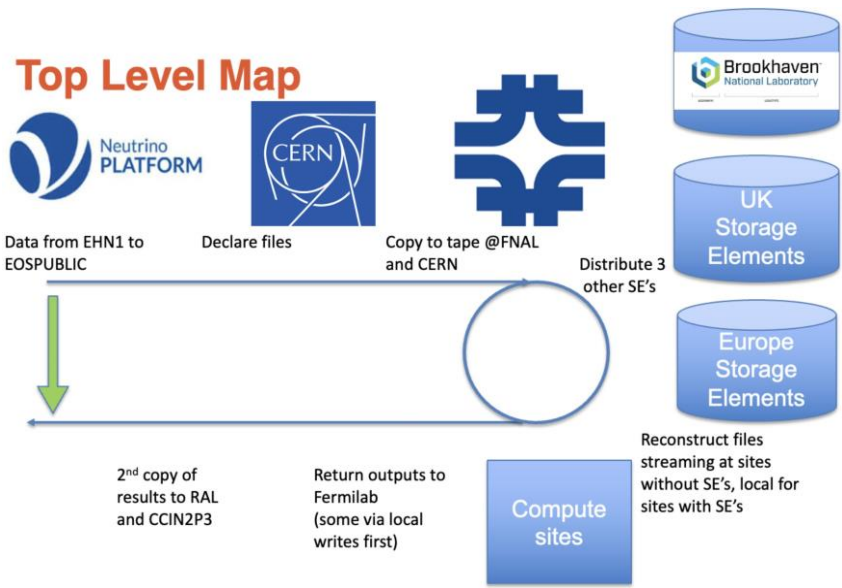

**Figure 3.** Raw data path for ProtoDUNE HD/VD raw data that was testing during the summer 2022 DUNE Data Challenge. A throughput of greater than 3.6 GBytes/s was achieved across the Atlantic ocean and into storage elements at FNAL. Additionally, automated replica creation was handled through Rucio policies at storage elements in North America and Europe.

Rucio transfers between storage elements are planned to be accomplished using token authentication. The completion of the WLCG DC 2024 will then lead into a time of significant development and commissioning of software and infrastructure prior to the operations of the first DUNE far detector module in 2029. This development will include the improvements in reconstruction algorithms[6], utilization of GPUs and accelerators for processing[7], and parallel processing approaches for optimized resource utilization[8].

## 6 Summary

The DUNE Offline Computing continues to operate successfully in support of both prototype detectors and the DUNE physics program as design and preparation for the construction of the DUNE detectors continues. The DUNE Computing Consortium has successfully developed a computing model that projects computing resources needs throughout the lifespan of DUNE operation, and the collaboration has begun contributing resources based upon that model. As well, DUNE Offline Computing is actively participating in data challenges both internally and through the WLCG to assure that the available computing resources are fully integrated and meet the needs of the experiment. During the next 10 years, considerable software design and development will take place in response to the unique needs of DUNE and in order to take advantage of new computing architectures. Additionally, DUNE Computing Consortium is working to develop new analysis centers that will provide easily accessible and powerful resources to physicists at the analysis stage.

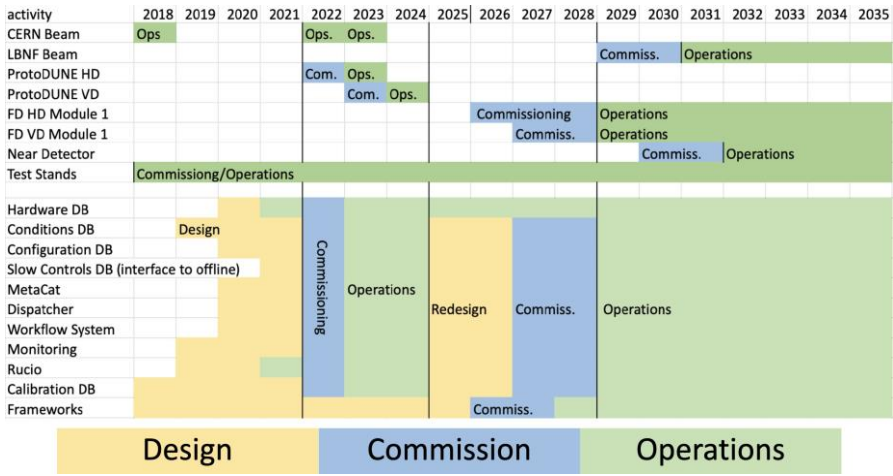

**Figure 4.** The proposed timeline for DUNE Offline Computing development and the associated experimental activities that motivate this development. At the time of publication of proceedings, the timeline for ProtoDUNE HD and VD operations has been postponed until 2024.

## 7 Acknowledgements

This document was prepared by the DUNE collaboration using the resources of the Fermi National Accelerator Laboratory (Fermilab), a U.S. Department of Energy, Office of Science, HEP User Facility. Fermilab is managed by Fermi Research Alliance, LLC (FRA), acting under Contract No. DE-AC02-07CH11359. This work was supported by CNPq, FAPERJ, FAPEG and FAPESP, Brazil; CFI, IPP and NSERC, Canada; CERN; MŠMT, Czech Republic; ERDF, H2020-EU and MSCA, European Union; CNRS/IN2P3 and CEA, France; INFN, Italy; FCT, Portugal; NRF, South Korea; CAM, Fundación "La Caixa", Junta de Andalucía-FEDER, MICINN, and Xunta de Galicia, Spain; SERI and SNSF, Switzerland; TÜBİTAK, Turkey; The Royal Society and UKRI/STFC, United Kingdom; DOE and NSF, United States of America.

The ProtoDUNE-SP and ProtoDUNE-DP detectors were constructed and operated on the CERN Neutrino Platform. We gratefully acknowledge the support of the CERN management, and the CERN EP, BE, TE, EN and IT Departments for NP04/ProtoDUNE-SP.

This research used resources of the National Energy Research Scientific Computing Center (NERSC), a U.S. Department of Energy Office of Science User Facility operated under Contract No. DE-AC02-05CH11231. Thanks to the Computation Sciences and Artificial Intelligence Directorate at Fermilab for operation of storage elements and services, development of new services, and consulting on operations. Thanks to the CERN IT staff for FTS3, EOS, and CTA support. Thanks to Oli Gutsche for numerous discussions about offline computing operations and organization.